\documentclass[aps,prl,twocolumn,superscriptaddress]{revtex4-2}

\usepackage{graphicx}
\usepackage{amsmath}
\usepackage{xcolor}

\begin{document}

\title{Triplon Bose-Einstein condensation and proximate magnetism in dimerized antiferromagnets}

\author{Z. Y. Zhao}
\affiliation{State Key Laboratory of Functional Crystals and Devices, Fujian Institute of Research on the Structure of Matter, Chinese Academy of Sciences, Fuzhou 350108, China}
\affiliation{State Key Laboratory of Structural Chemistry, Fujian Institute of Research on the Structure of Matter, Chinese Academy of Sciences, Fuzhou, Fujian 350002, China}

\author{F. Y. Li}
\affiliation{State Key Laboratory of Surface Physics and Department of Physics, Fudan University, Shanghai 200433, China}

\author{C. Dong}
\affiliation{Institute for Solid State Physics, University of Tokyo, Kashiwa, Chiba 277-8581, Japan}
\affiliation{Wuhan National High Magnetic Field Center, Huazhong University of Science and Technology, Wuhan, Hubei 430074, China}

\author{R. Chen}
\affiliation{Wuhan National High Magnetic Field Center, Huazhong University of Science and Technology, Wuhan, Hubei 430074, China}

\author{M. Y. Cui}
\affiliation{State Key Laboratory of Functional Crystals and Devices, Fujian Institute of Research on the Structure of Matter, Chinese Academy of Sciences, Fuzhou 350108, China}

\author{Z. W. Ouyang}
\affiliation{Wuhan National High Magnetic Field Center, Huazhong University of Science and Technology, Wuhan, Hubei 430074, China}

\author{J. F. Wang}
\affiliation{Wuhan National High Magnetic Field Center, Huazhong University of Science and Technology, Wuhan, Hubei 430074, China}

\author{Y. Kohama}
\email{ykohama@issp.u-tokyo.ac.jp}
\affiliation{Institute for Solid State Physics, University of Tokyo, Kashiwa, Chiba 277-8581, Japan}

\author{Z. Z. He}
\email{hezz@fjirsm.ac.cn}
\affiliation{State Key Laboratory of Functional Crystals and Devices, Fujian Institute of Research on the Structure of Matter, Chinese Academy of Sciences, Fuzhou 350108, China}

\author{Gang v.~Chen}
\email{chenxray@pku.edu.cn}
\affiliation{International Center for Quantum Materials, School of Physics, Peking University, Beijing 100871, China}
\affiliation{Collaborative Innovation Center of Quantum Matter, 100871 Beijing, China}

\date{\today}

\begin{abstract}

Dimerized quantum magnets provide a useful arena for novel quantum states and phases transitions with the singlet-triplet type of triplon excitations. Here we study the triplon physics and the Bose-Einstein condensation in two isostructural dimerized antiferromagnets $A$Cu(SeO$_3$)$_2$ ($A$ = Hg, Cd). With the systematic measurements, we demonstrate a dimer singlet ground state in HgCu(SeO$_3$)$_2$ with a triplon gap $\sim$ 7.9 K and a triplon Bose-Einstein condensation with an antiferromagnetic order in CdCu(SeO$_3$)$_2$ below 4.4 K. We further adopt the bond-operator technique and show that the elemental replacement preserves the Hamiltonian and allows the study in a unified theoretical framework with tunable interdimer and intradimer interactions
on the opposite sides of the quantum critical point. With the peculiar Cu$_2$O$_8$ dimer configuration and effective ferromagnetic interdimer interaction, $A$Cu(SeO$_3$)$_2$ is distinguished from other $S$ = 1/2 dimerized antiferromagnets. Our results represent a global understanding of the magnetic ground states as well as the magnetic transitions in the dimerized magnets of this unusual crystal structure.

\end{abstract}

\maketitle

\begin{figure}
\centering
\includegraphics[clip,width=8.5cm]{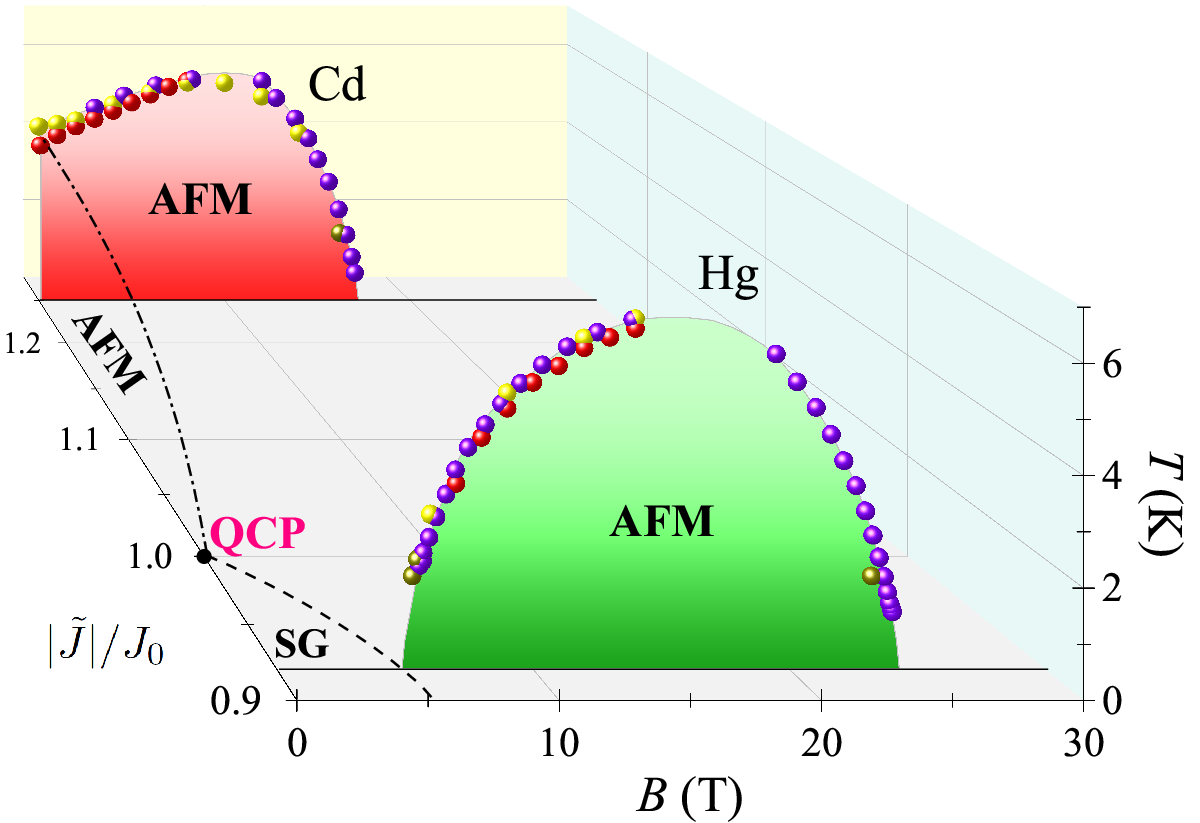}
\caption{${\mid}\tilde{J}{\mid}$/$J_0$-$B$-$T$ phase diagrams for $A$Cu(SeO$_3$)$_2$ ($A$ = Hg, Cd) determined from magnetic susceptibility (red), specific heat (yellow), high-field magnetization (brown), and magnetocaloric (purple) results. The ratio ${\mid}\tilde{J}{\mid}$/$J_0$ is a decisive parameter for the magnetic ground state with the quantum critical point (QCP) lying at ${\mid}\tilde{J}{\mid}$/$J_0$ = 1. $B$-$T$ phase diagrams locate at ${\mid}\tilde{J}{\mid}$/$J_0$ = 0.92 for Hg and 1.25 for Cd compounds. Enclosed regimes by the dashed line in the ${\mid}\tilde{J}{\mid}$/$J_0$-$B$ plane and dashed-dotted line in the ${\mid}\tilde{J}{\mid}$/$J_0$-$T$ plane outline the spin-gapped (SG) and antiferromagnetic (AFM) phases, respectively.}
\end{figure}

Multiflavor Mott insulators with the singlet-triplet energy level structure represent a characteristic feature for the emergence of the triplon or triplon-like excitations and their physics in quantum magnets \cite{npj-9-1,RMP-86-563,PR-1070-1}. Such an energy level structure could appear in, for example, dimerized antiferromagnets \cite{RMP-86-563}, $4d^4/5d^4$ spin-orbit-coupled Mott insulators with octahedral crystal environments \cite{PRL-111-197201,PRB-91-054412}, $3d^6/3d^8$ spin-orbit-coupled Mott insulators with tetrahedral crystal environments \cite{PRB-100-045103,PRL-102-096406}, rare-earth magnets \cite{PRB-99-224407,PRB-109-094423}, cluster Mott insulators \cite{PRB-93-245134,PRB-97-035124,PRL-111-217201}, et al.
For instance, the local ground state of the magnetic ion in the relevant spin-orbit-coupled Mott insulators with this energy level structure is a spin-orbital singlet, and the effective triplon is a spin-orbital exciton. In the dimerized antiferromagnet \cite{RMP-86-563} that is probably a more stereotype system with this singlet-triplet structure, the ground state of the strongly antiferromagnetically coupled bond is a real spin singlet, and the resulting excited states or triplon modes are the spin triplet states of this bond. The interaction and the dynamics of these triplons play a key role in the
understanding of the emergent many-body physics in these multiflavor Mott insulators.

Here we work with two isostructural dimerized quantum antiferromagnetic materials, $A$Cu(SeO$_3$)$_2$ ($A$ = Hg, Cd), to reveal the emergent triplon physics and the proximate magnetism from the triplon Bose-Einstein condensation. These two materials are located on the opposite side of the triplon condensation transition and thus provide an ideal platform to study the distinct magnetic properties on both sides of the quantum phase transition. In the often-referenced Shastry-Sutherland lattice antiferromagnet SrCu$_2$(BO$_3$)$_2$, the intradimer exchange interaction is predominant at ambient pressure \cite{PRL-111-137204}, placing the compound in the dimer singlet phase \cite{PRL-133-246702} with a large triplon gap \cite{NP-13-962}. The hydrostatic pressure has been used to induce a plaquette-singlet state and a N\'{e}el order by varying
the interactions \cite{NC-7-11956,NP-13-962,Nature-592-370,NC-13-2301,Science-380-1179},
and the transition between these two states are likely to be a deconfined quantum criticality \cite{Science-380-1179}. The chemical substitution both in and out of the Shastry-Sutherland layers has been claimed to break some Cu$^{2+}$ dimers that does not drive a
triplon condensation~\cite{PRB-71-014441,JCG-306-123,PRB-76-214427,PRL-97-247206,NC-10-2439,IC-63-20335,PRB-73-014414}.
So far, the chemical-pressure-induced phase transition via elemental replacement or doping has been scarcely discovered in dimerized magnets~\cite{RMP-86-563}. In this Letter, we point out that $A$Cu(SeO$_3$)$_2$ could serve as an exceptional object to capture this issue, and allow us to access the global phase diagram of the triplon Bose-Einstein condensation with both magnetic fields and chemical pressure [Fig. 1].

$A$Cu(SeO$_3$)$_2$ crystallizes in a monoclinic structure \cite{DT-44-11420}, and equivalent Cu$^{2+}$ ions occupy a distorted square pyramid geometry with a symmetry related O4$^i$ at the vertex [Fig. 2(a)]. Two CuO$_5$ square pyramids separated by $d_0$ constitute a centrosymmetric Cu$_2$O$_8$ dimer by sharing the O4-O4$^i$ edge, giving a net vanishing Dzyaloshinskii-Moriya interaction \cite{JPCS-4-241, PR-120-91}. Each dimer links four neighbours through a single SeO$_3$ scaffold in the $bc$ plane ($d_1$, $d_2$) and is double-bridged to adjacent entities along the $a$ axis ($d_3$) [Figs. 2(b, c)]. The refined crystallographic data of our single crystals are given in Supplemental Material \cite{SM}. The prior study, that proposed a robust gapped ground state for CdCu(SeO$_3$)$_2$ \cite{DT-44-11420}, is quite questionable. Observations of the finite magnetic susceptibility approaching zero kelvin, nonzero initial magnetization, as well as the incipient anomaly in the specific heat definitely point to a magnetically ordered ground state. Thus, we revisit the magnetic properties of $A$Cu(SeO$_3$)$_2$ on a closer inspection, and their phase diagrams are depicted in Fig. 1. A magnetic order is explicitly recognized in CdCu(SeO$_3$)$_2$, and a drastic decrease of the intradimer coupling $J_0$ is illuminated to be the primary force to drive the triplon gap to zero. The unique traits including the peculiar Cu$_2$O$_8$ configuration and effective ferromagnetic interdimer coupling $J'$ are discussed to be essential to distinguish $A$Cu(SeO$_3$)$_2$ from other dimerized magnets.

\begin{figure}
\centering
\includegraphics[clip,width=8.5cm]{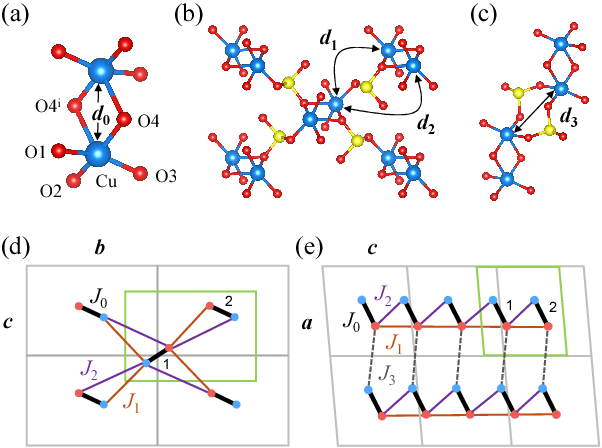}
\caption{(a-c) Multiple exchange pathways. $d_0$ is the nearest distance within a dimer, $d_1$, $d_2$, and $d_3$ are the interdimer distances with different connections to the SeO$_3$ trigonal pyramids. (d, e) The dimer network projections in the $bc$ and $ac$ planes. Cu$^{2+}$ ions depicted by pink (cyan) balls are referred as ``left" (``right") in the theoretical approach~\cite{SM}.
The thick black bond denotes $J_0$, while the thin bonds denoted $J_1$ (pink), $J_2$ (purple),
and $J_3$ are labelled by the grey dashed line. There are two dimers in each unit cell, labelled by ``1" and ``2". For convenience, a different unit cell (green boxes) rather than the standard one (gray boxes) is chosen in the theoretical approach~\cite{SM}.}
\end{figure}

Comparative results of specific heat $C\rm_p$, magnetic susceptibility $\chi$, and high-field magnetization $M$ demonstrate contrastive magnetism relying on the $A$-site cation [Figs. 3(a-c)]. A gapped dimer singlet state is confirmed in HgCu(SeO$_3$)$_2$ as evidenced by the absence of an anomaly in $C\rm_p$ down to 0.4 K and the rapid drop of $\chi\rm_{Hg}$ at lower temperatures. At $T$ = 1.7 K, $M$($B$) remains zero until the triplon gap is closed at the field-driven triplon condensation with the critical field $B\rm_{c1}$ $\sim$ 5.3 T, and then monotonously increases to a saturation about 1.16 $\mu\rm_B$ above $B\rm_{c2}$ $\sim$ 23.2 T. The zero-field triplon gap is estimated to be $\Delta$ = $g\mu{\rm_B}B\rm_{c1}$ = 7.9 K using $g$ = 2.21 obtained from ESR data \cite{SM}. On the contrary, a magnetic order is present in CdCu(SeO$_3$)$_2$ by the emergence of a weak but unambiguous anomaly at $T\rm_{N,Cd}$ = 4.4 K in the zero-field $C\rm_p$. Additionally, a much larger $\chi\rm_{Cd}$ and finite initial slope of $M$($B$) approve the occurrence of a magnetic order. The moment monotonously increases as the field is swept up,
and the Cu$^{2+}$ spins are completely polarized above $B\rm_s$ $\sim$ 16.2 T. Thus, the different magnetism in $A$Cu(SeO$_3$)$_2$ implies that the triplon gap could be directly suppressed at ambient pressure simply by replacing Hg by Cd. Equivalently, a chemical pressure-induced triplon condensation and the proximate magnetism can be realized in $A$Cu(SeO$_3$)$_2$.

\begin{figure*}
\centering
\includegraphics[clip,width=17.5cm]{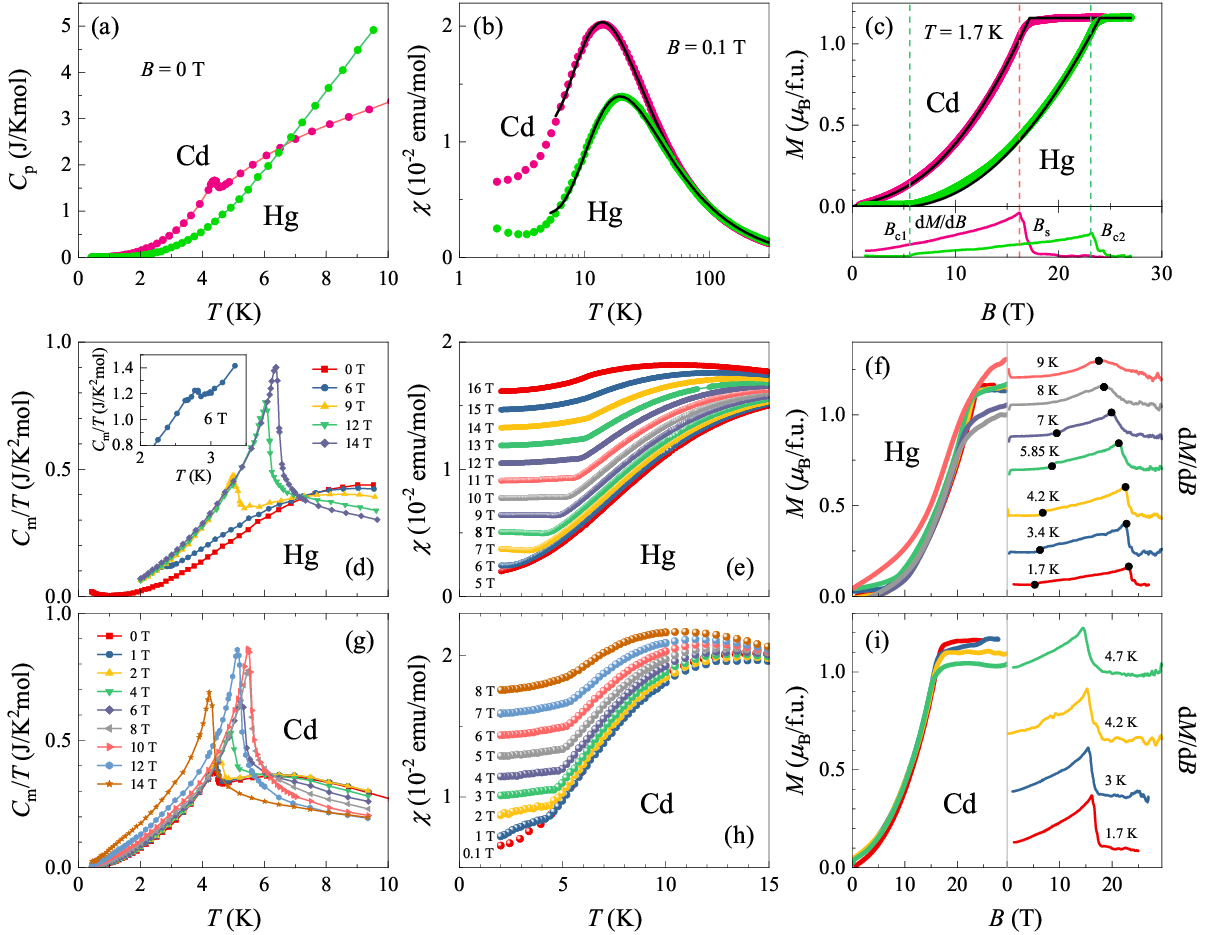}
\caption{(a-c) Comparative results of zero-field specific heat, low-field magnetic susceptibility, and high-field magnetization at 1.7 K. Solid lines in (b) are the fittings using mean-field modified Bleaney-Bowers model above 5 K, which yields $J_0$ = 32.16(9) K, $J'$ = -9(2) K, $g$ = 2.33 for HgCu(SeO$_3$)$_2$, and $J_0$ = 24.44(7) K, $J'$ = -17.7(6) K, $g$ = 2.21 for CdCu(SeO$_3$)$_2$. The negative $J'$ indicates an effective ferromagnetic interdimer interaction. Solid lines in (c) are the theoretical zero-temperature longitudinal magnetization. $B\rm_{c1}$ and $B\rm_{c2}$ are the lower and upper critical fields for HgCu(SeO$_3$)$_2$, while $B\rm_s$ is the critical field for CdCu(SeO$_3$)$_2$, which are determined from the derivatives in the lower panel. (d-i) Magnetic specific heat, magnetic susceptibility, and high-field magnetization along with their derivatives for Hg and Cd compound, respectively, in higher fields and at higher temperatures. Inset in (d) is a zoom in of the 6-T curve around the transition. Black dots in the right panel of (f) are the guides for the transition fields.}
\end{figure*}

To have an in-depth investigation of the magnetic properties, more detailed results are presented in Figs. 3(d-i). The magnetic specific heat $C\rm_m$ has been separated from the lattice contributions \cite{SM}. In HgCu(SeO$_3$)$_2$, a weak hump starts to appear in $C\rm_{m,Hg}$ around $T\rm_{N,Hg}$ = 2.8 K in 6 T, which evolves into a $\lambda$ peak at higher temperatures with a growing magnitude in higher fields [Fig. 3(d)]. Meanwhile, $\chi\rm_{Hg}$ exhibits a weak slope change at 3.4 K in 7 T and turns into a plateau below $T\rm_{N,Hg}$ in higher fields [Fig. 3(e)]. On the other hand, in CdCu(SeO$_3$)$_2$, the $C\rm_{m,Cd}$ anomaly are slightly changed below 2 T but are strengthened in position and magnitude rapidly above 4 T [Fig. 3(g)]. $T\rm_{N,Cd}$ achieves to 5.5 K in 8 T and after that the $\lambda$ peak is weakened gradually instead. $\chi\rm_{Cd}$ shows a tiny slope change already in 0.1 T, which converts progressively to a linear temperature dependence in higher fields [Fig. 3(h)]. As shown in Figs. 3(f), 3(i) \cite{SM}, the boundaries of the antiferromagnetic (AFM) phase deduced from the high-field magnetization are much higher than those obtained in static fields for both compounds due to the strong magnetocaloric effect (MCE). To accurately depict the magnetic field versus temperature phase diagrams of Fig. 1, MCE describing the temperature change of magnetic materials in response to the external magnetic field under the quasiadiabatic conditions is therefore an advantageous probe \cite{RSI-81-104902,RSI-84-074901}.

\begin{figure*}
\centering
\includegraphics[clip,width=17.5cm]{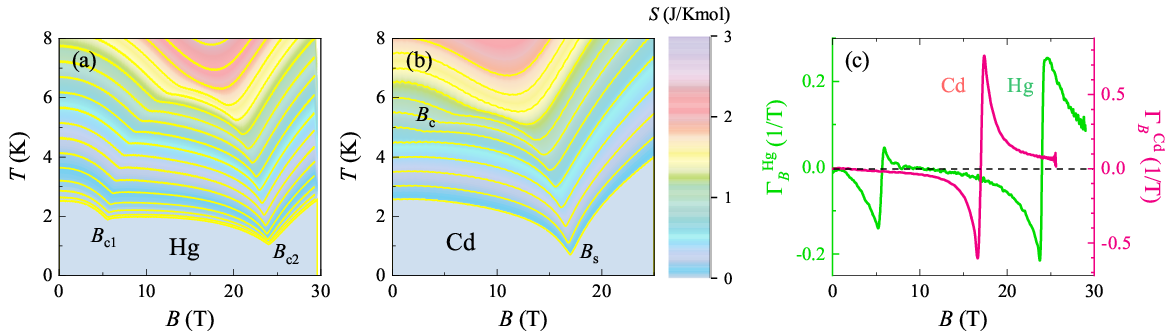}
\caption{(a, b) Contour maps of $S$($T$, $B$) for Hg and Cd compound, respectively. (c) Magnetic Gr\"{u}neisen parameter $\Gamma_B$. The data are collected in the field-down measurements.}
\end{figure*}

In the vicinity of a field-induced quantum critical point (QCP), the low-temperature magnetic properties are governed by strong quantum fluctuations that give rise to a peculiar entropy landscape.
The spin entropy is accumulated near the phase boundary as a consequence of the competition between different ground states separated by the QCP, leading to a minimum of the isentropes \cite{IJMPB-28-1430017}. Figures 4(a, b) plot the isentropic contours as a function of temperature and magnetic field in both compounds. For HgCu(SeO$_3$)$_2$, with increasing field at the lowest temperature, the spin entropy is significantly accumulated due to the gradual suppression of the triplon gap. As a result, the sample temperature is firstly lowered and develops a minimum at $B\rm_{c1} \sim$ 5.6 T where the triplon condensation occurs. With further increasing field, the Cu$^{2+}$ spins are arranged in an AFM order until $B\rm_{c2} \sim$ 24.0 T above which the spins are fully polarized and the sample temperature is rapidly increased. In CdCu(SeO$_3$)$_2$ where the triplons have already been condensed at zero field, at the lowest temperature, the sample temperature is monotonously decreased and reaches a minimum at $B\rm_s$ $\sim$ 17.0 T above which the spins are fully polarized. This part of the behaviour can be compared in parallel with the Hg compound in the regime above $B\rm_{c1}$ where the triplons are condensed, and the behaviours are quite similar qualitatively. At 5 K where the thermal fluctuation suppresses the magnetic order and triplon condensation at zero field, a second minimum starts to appear at $B\rm_c$ $\sim$ 4.6 T with a rather weak magnetocaloric response. $T$($B$) curves always exhibit minimum feature at the critical fields, and the distance between them is narrowed upon warming. Their minimum positions outline the phase boundaries of the AFM order, which coincide with the anomalies in $\chi$($T$), $C\rm_p$, and $M$($B$) measurements, see Fig. 1. Based on the MCE data, the magnetic Gr\"{u}neisen parameter $\Gamma_B$ = (1/$T$)(${\partial}T$/${\partial}B$)$_S$ can be derived. As shown in Fig. 4(c), $\Gamma_B$ is divergent and undergoes a sign change when crossing the quantum phase transitions \cite{PRL-91-066404,PRB-72-205129}.

Realization of different ground states between the Hg and Cd isomorphs essentially clarifies the existence of a chemical pressure-induced QCP in Hg$_x$Cd$_{1-x}$Cu(SeO$_3$)$_2$. According to the bond-operator mean-field approximation \cite{PRB-66-104405}, the triplon gap of a three-dimensional dimerized magnet can be expressed as $\Delta$ = $g\mu{\rm_B} B\rm_{c1}$ = $J_0 {(1-{\mid}\tilde{J}{\mid}/J_0})^{\frac{1}{2}}$, where ${\mid}\tilde{J}{\mid}$ is a linear combination of the interdimer interactions per dimer. The ratio ${\mid}\tilde{J}{\mid}$/$J_0$ reflects
the tendency to close the triplon gap and can be used as a measure to characterize the distance
between the system and the QCP. The zero-field QCP is located at ${\mid}\tilde{J}{\mid}$/$J_0$ = 1 and separates the gapped dimer singlet state (${\mid}\tilde{J}{\mid}$/$J_0$ $<$ 1) and the AFM state from the triplon condensation (${\mid}\tilde{J}{\mid}$/$J_0$ $>$ 1) [Fig. 1].

As illustrated in Figs. 2(d, e), there are four symmetry-equivalent Cu$^{2+}$ ions in each crystal unit cell and can be grouped into two dimers with strong AFM intradimer coupling $J_0$. Three kinds of interdimer couplings $J_1$, $J_2$, $J_3$ (corresponding to $d\rm_{1,2,3}$) are further considered to construct a minimal model \cite{SM}. Via a mean-field theory, the physical quantities like magnetization and $T\rm_N$ can be calculated \cite{SM}. In $A$Cu(SeO$_3$)$_2$, ${\mid}\tilde{J}{\mid}$ $\equiv$ 2$\mid$$J_1-J_2$$\mid$ + $\mid$$J_3$$\mid$ and $J'$ $\equiv$ 2($J_1$ + $J_2$) + $J_3$. Fitting to the $M$($B$) curves [solid lines in Fig. 3(c)] gives $J_0$ = 32.6 K, ${\mid}\tilde{J}{\mid}$ = 30.1 K, $J'$ = -20.6 K, $g$ = 2.32 for HgCu(SeO$_3$)$_2$, and $J_0$ = 22.0 K, ${\mid}\tilde{J}{\mid}$ = 27.6 K, $J'$ = -18.0 K, $g$ = 2.32 for CdCu(SeO$_3$)$_2$. Thus, ${\mid}\tilde{J}{\mid}$/$J_0$ is calculated to be 0.92 and 1.25, respectively, positioning the former in the gapped dimer singlet phase (very close to the QCP)
and the latter in the AFM phase [Fig. 1]. We find that replacement of Hg by Cd results in a remarkable decrease of $J_0$, strongly pushing the AFM region towards the lower fields and ultimately prevails
over the dimer singlet state in zero field. Moreover, the negative $J'$ in both compounds signals an effective ferromagnetic interdimer interaction, which is rarely found in the dimerized magnets \cite{PRL-112-207201,PRB-95-024404}. The well-reproduced $\chi$($T$) [solid lines in Fig. 3(b)] using the mean-field modified Bleaney-Bowers expression \cite{JPSJ-75-054706} and the rather weak Weiss temperature \cite{SM} consistently supports the presence of effective ferromagnetic $J'$.

Unlike other Cu$^{2+}$-based dimerized magnets \cite{PRB-78-224403,JPSJ-75-064703,PRB-93-174121}, $J_0$ in $A$Cu(SeO$_3$)$_2$ is much weaker and can be rationally understood from the crystal structure. It is known that for a regular CuO$_5$ square pyramid with Cu$^{2+}$ ion lying in the center of the basal plane, only $d_{x^2-y^2}$ orbital is active and no overlap is expected in a Cu$_2$O$_8$ dimer with two parallelized basal planes connected by a side edge, like the situation of this work. As the Cu$^{2+}$ ion is pulled inside the pyramid by a distance $\delta z$, the energy difference between $d_{x^2-y^2}$ and $d_{3z^2-r^2}$ levels is reduced, and the level sequence could even be inverted if the shift is large enough \cite{Book-orbital}. This is probably responsible for the presence of weak $J_0$ in $A$Cu(SeO$_3$)$_2$ due to a superposition of $d_{x^2-y^2}$ and $d_{3z^2-r^2}$ orbitals, which is no doubt subtle to the modification of local distortion. In light of the small difference of the ionic radii between Hg$^{2+}$ and Cd$^{2+}$ cations (1.14 {\AA} \textit{vs} 1.10 {\AA}) \cite{IR}, the Cu-O bond lengths as well as the intradimer distance are slightly altered \cite{SM}. The pronounced change of $J_0$, therefore, cannot be ascribed to the size effect. The stronger electrostatic attraction of Hg to O4$^i$ atom contributes to a weaker twist of CuO$_5$ square pyramid, and the larger $J_0$ in HgCu(SeO$_3$)$_2$ is consequently birthed from the resultant enhanced orbital mixture \cite{SM}.

In the dimerized magnets, the domelike phase diagram with a cusplike feature in $\chi$($T$) is predicted to meet a relation $B\rm_{c1}$($T$)-$B\rm_{c1}$(0) $\propto$ $T\rm^{3/2}$ \cite{PRL-84-5868} from the picture of triplon Bose-Einstein condensation (BEC) \cite{RMP-86-563}. In HgCu(SeO$_3$)$_2$, the onset temperature of the field-induced magnetic order also follows a power law but yields a much larger critical exponent $\phi$ = 2.2(2) \cite{SM}. The overestimated $\phi$ is probably due to the higher temperature window since the $T\rm^{3/2}$ relation is established by assuming a dilute boson limit and is only valid at sufficiently low temperatures, e.g. $T \leq$ 0.4$T\rm_{N,max}$ ($\sim$ 2.6 K) \cite{RMP-86-563}. One possibility is the violation of the dilute triplon condition \cite{PRB-66-104405}. In Fig. 3(e), $\chi\rm_{Hg}$ exhibits a weak cusplike minimum at $T\rm_N$ when $B \sim B\rm_{c1}$, but becomes nearly constant below $T\rm_N$ in higher fields. In the BEC theory of dilute triplons \cite{PRL-84-5868}, the cusplike minimum at $T\rm_N$ is due to the net increase of triplons considering that the amount of condensed triplons is greater than that of thermally excited uncondensed triplons. Nevertheless, with increasing fields, the density of condensed triplons becomes so large that the condition of dilute triplons is no longer satisfied \cite{PRB-66-104405}. In the dense triplon region, the mean-field approach predicting a constant magnetization below $T\rm_N$ describes better the transition \cite{JPSJ-28-1413}. The weak temperature dependence of $\chi\rm_{Hg}$ thus reflects a high population of condensed triplons, which is demonstrated by the larger critical density $n\rm_{cr}$ \cite{SM}. This is plausible since HgCu(SeO$_3$)$_2$ is close to the QCP such that the triplon condensation is unconventionally accelerated. When the triplon density exceeds a threshold, a magnetic order eventually emerges. This is what happens in CdCu(SeO$_3$)$_2$ in view of the same temperature dependencies of $C\rm_{m,Cd}$(0 T) and $C\rm_{m,Hg}$(14 T) \cite{SM} and the similar flat $\chi\rm_{Cd}$ below $T\rm_N$ [Fig. 3(h)].

\emph{Discussion.}---From the experimental and theoretical analysis, $A$Cu(SeO$_3$)$_2$ is found to be a novel $S$ = 1/2 dimerized series to host a chemical pressure-induced triplon BEC and the proximate magnetic order. It is of interest to draw a comparison with another set of dimerized magnets $A$CuCl$_3$ ($A$ = Tl, NH$_4$). TlCuCl$_3$ is a well-studied triplon BEC candidate with ${\mid}\tilde{J}{\mid}$/$J_0$ $\sim$ 0.83 \cite{PRB-69-054423}, $B\rm_{c1}$ $\sim$ 5.3 T, and $B\rm_{c2}$ $\sim$ 86.1 T \cite{PRL-125-267207}. After replacing Tl by NH$_4$, the rotational dynamics of NH$_4$ complexes introduces two structural transitions and three nearly decoupled Cu$^{2+}$ dimer subsystems at lower temperatures \cite{PRL-93-037207}. In zero field, one subsystem magnetically orders at 1.28 K while the rest two subsystems remain in the dimer singlet state with different triplon gaps. In contrast, $A$Cu(SeO$_3$)$_2$ is unique and unusual. On the one hand, albeit having almost the same gap as TlCuCl$_3$, HgCu(SeO$_3$)$_2$ locates much closer to the QCP with ${\mid}\tilde{J}{\mid}$/$J_0 \sim$ 0.92. Moreover, HgCu(SeO$_3$)$_2$ has a much lower $B\rm_{c2}$ ($\sim$ 23.2 T) due to the effective ferromagnetic $J'$, considering $g\mu{\rm_B} B\rm_{c2}$ = $J_0$ + (${\mid}\tilde{J}{\mid}$ + $J'$)/2. On the other hand, different from NH$_4$CuCl$_3$, the magnetic order in CdCu(SeO$_3$)$_2$ is driven without experiencing a structural transition and the vanishing of the triplon gap is stemmed from the enlarged ${\mid}\tilde{J}{\mid}$/$J_0$.

The pecular up-down Cu$_2$O$_8$ dimer configuration plays a key role in distinguishing $A$Cu(SeO$_3$)$_2$ from other dimerized magnets. In (Tl, K)CuCl$_3$ \cite{JPSJ-75-064703, PRB-78-224403} and (Sr, Ba)$_3$Cr$_2$O$_8$ \cite{PRB-81-014415, JPSJ-75-054706}, due to the large ion-radius difference, the positive chemical pressure shortens $d_0$ dramatically and increases the gap accordingly. This is in sharp contrast to the invariant $d_0$ as well as weakened $J_0$ in $A$Cu(SeO$_3$)$_2$, in which a minor volume compression is powerful enough to overcome the gap and trigger a magnetic order.

To summarize, we have convincingly shown that $A$Cu(SeO$_3$)$_2$ series is an intriguing material platform to study the singlet-triplet type of triplon excitations, triplon BEC, and the QCP-related physics. Observations of the chemical composition-induced QCP in Hg$_x$Cd$_{1-x}$Cu(SeO$_3$)$_2$ and the magnetic order driven by hydrostatic pressure in HgCu(SeO$_3$)$_2$ deserve further explorations.

\textit{Acknowledgements}---This work is financed by the National Natural Science Foundation of China (Grants No. 22205234 and No. 22175173) and the Interdisciplinary Program of Wuhan National High Magnetic Field Center (Grant No. WHMFC202125), Huazhong University of Science and Technology. G.C. thanks for the support from the MOST of China with Grants No.~2021YFA1400300, and the Fundamental Research Funds for the Central Universities, Peking University. Y.K. acknowledges the support from JSPS KAKENHI Grant Number 25H00600.

\clearpage
\newpage

\begin{figure*}
\centering
\includegraphics[clip,width=17.5cm]{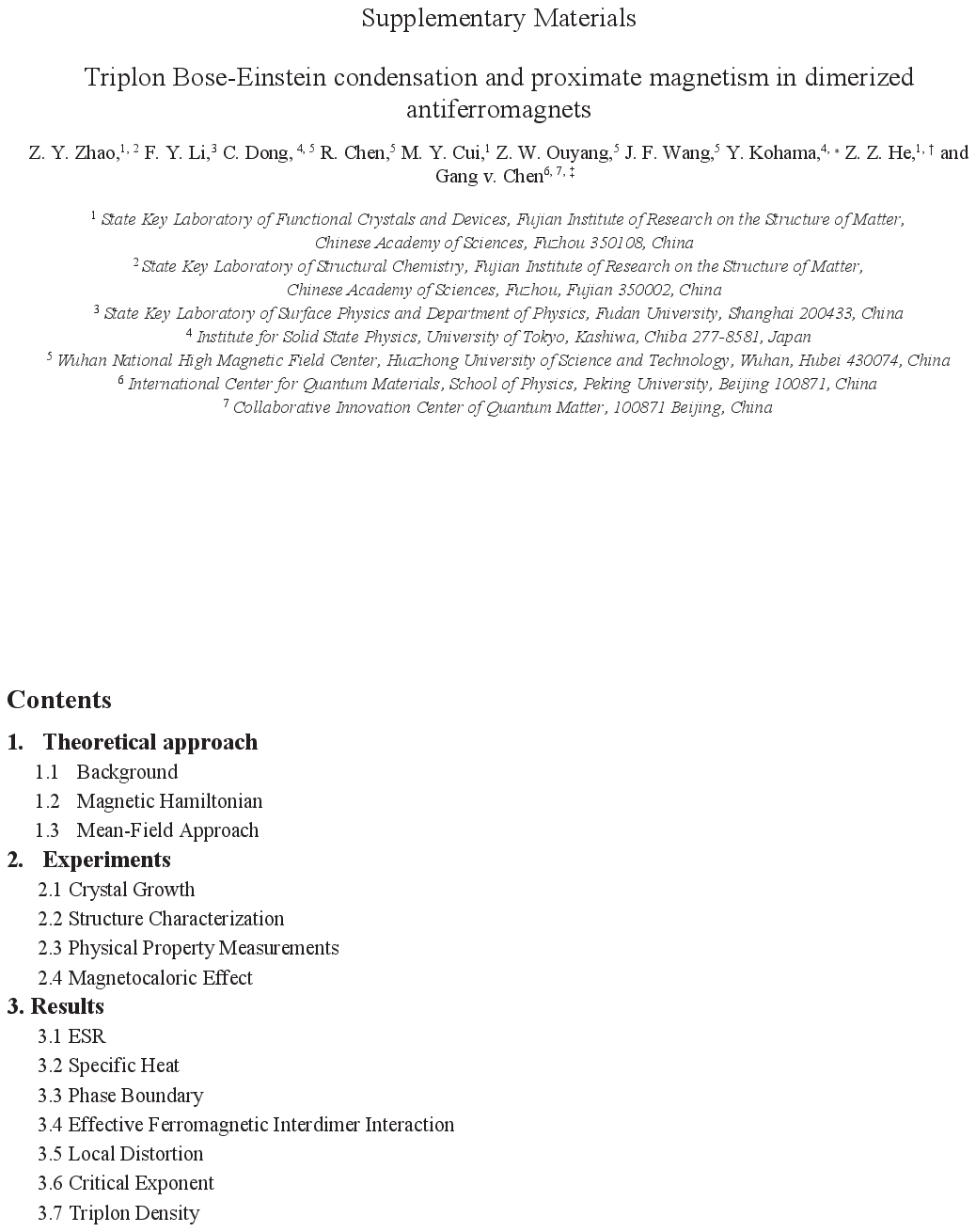}
\end{figure*}

\begin{figure*}
\centering
\includegraphics[clip,width=17.5cm]{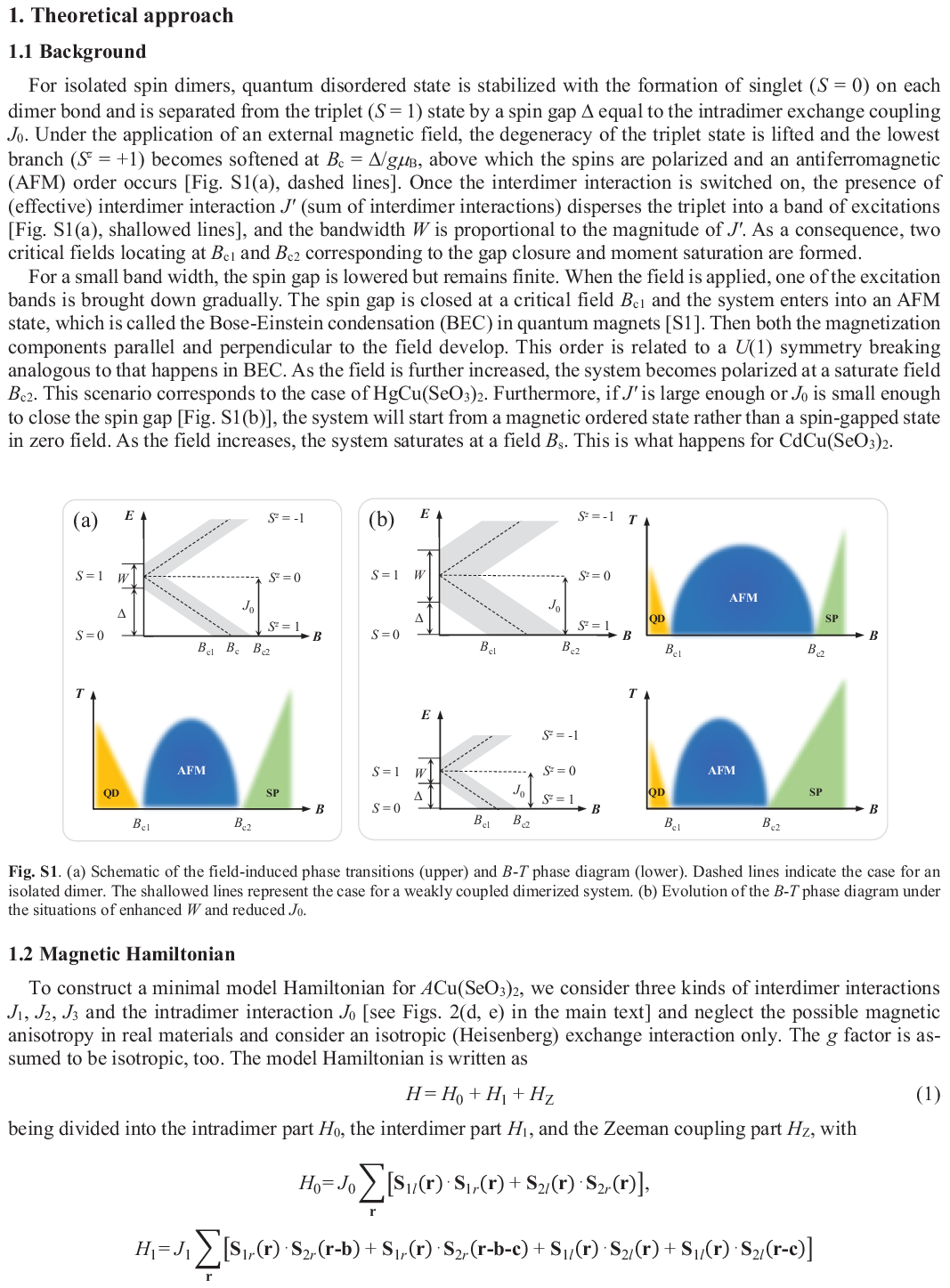}
\end{figure*}

\begin{figure*}
\centering
\includegraphics[clip,width=17.5cm]{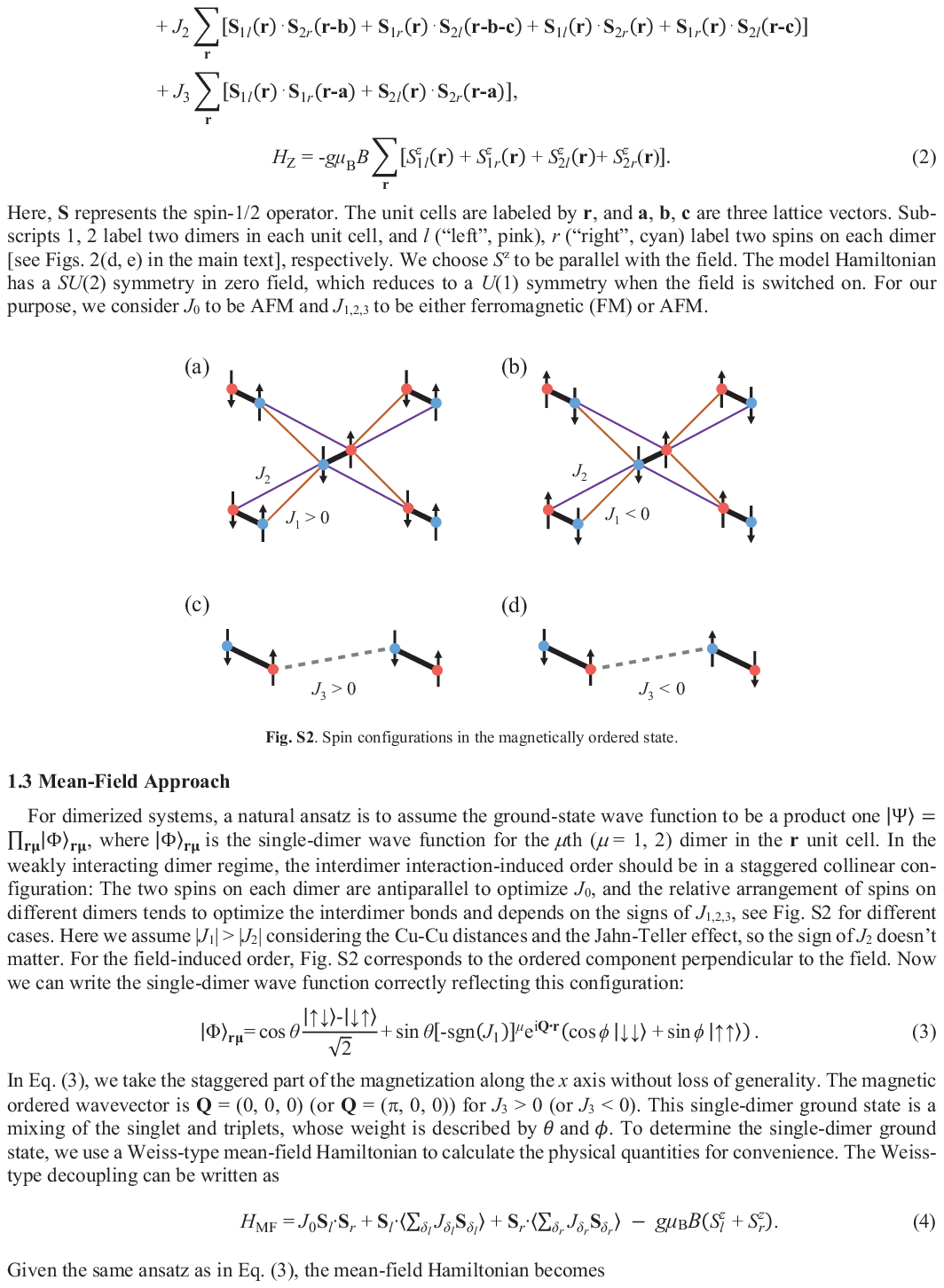}
\end{figure*}

\begin{figure*}
\centering
\includegraphics[clip,width=17.5cm]{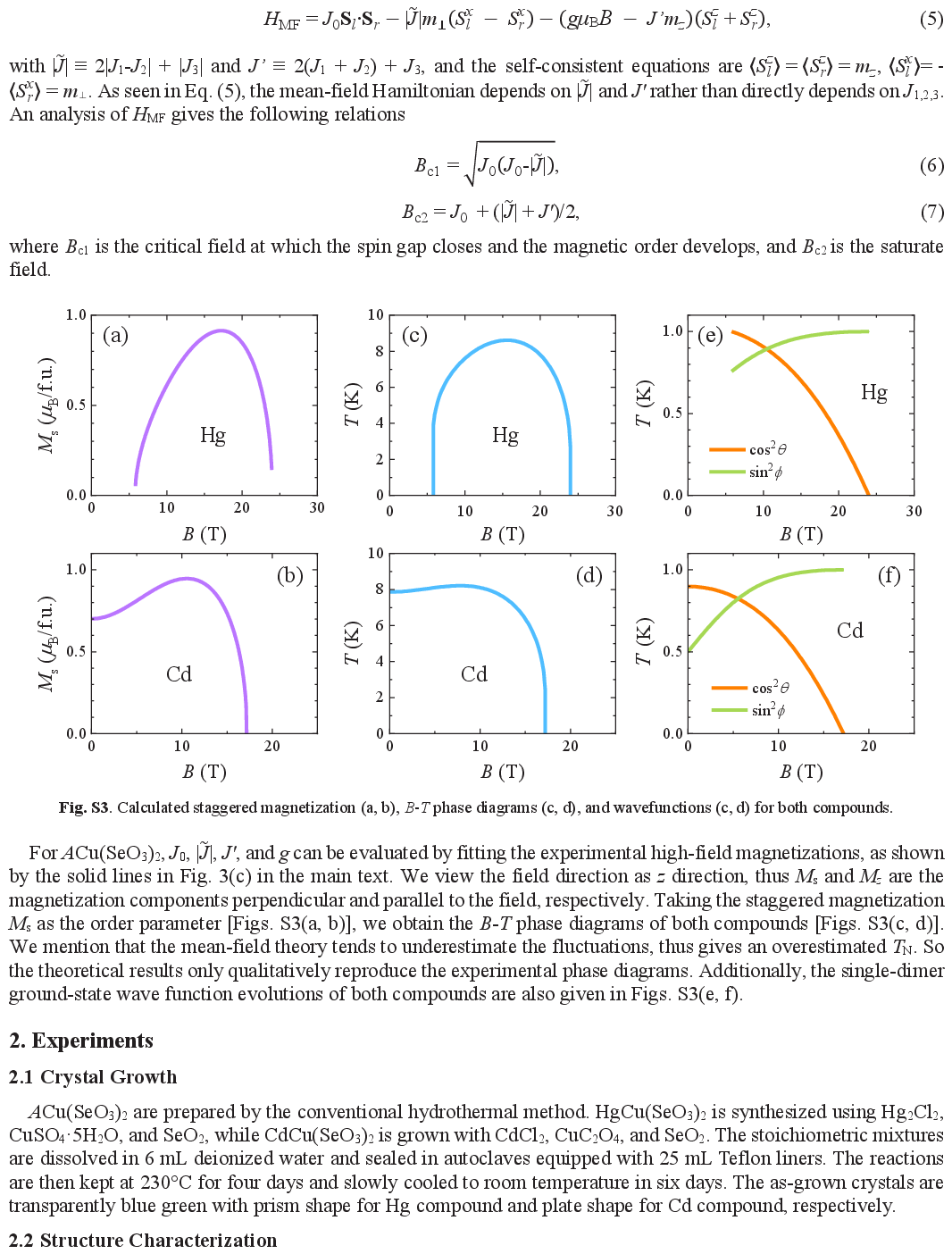}
\end{figure*}

\begin{figure*}
\centering
\includegraphics[clip,width=17.5cm]{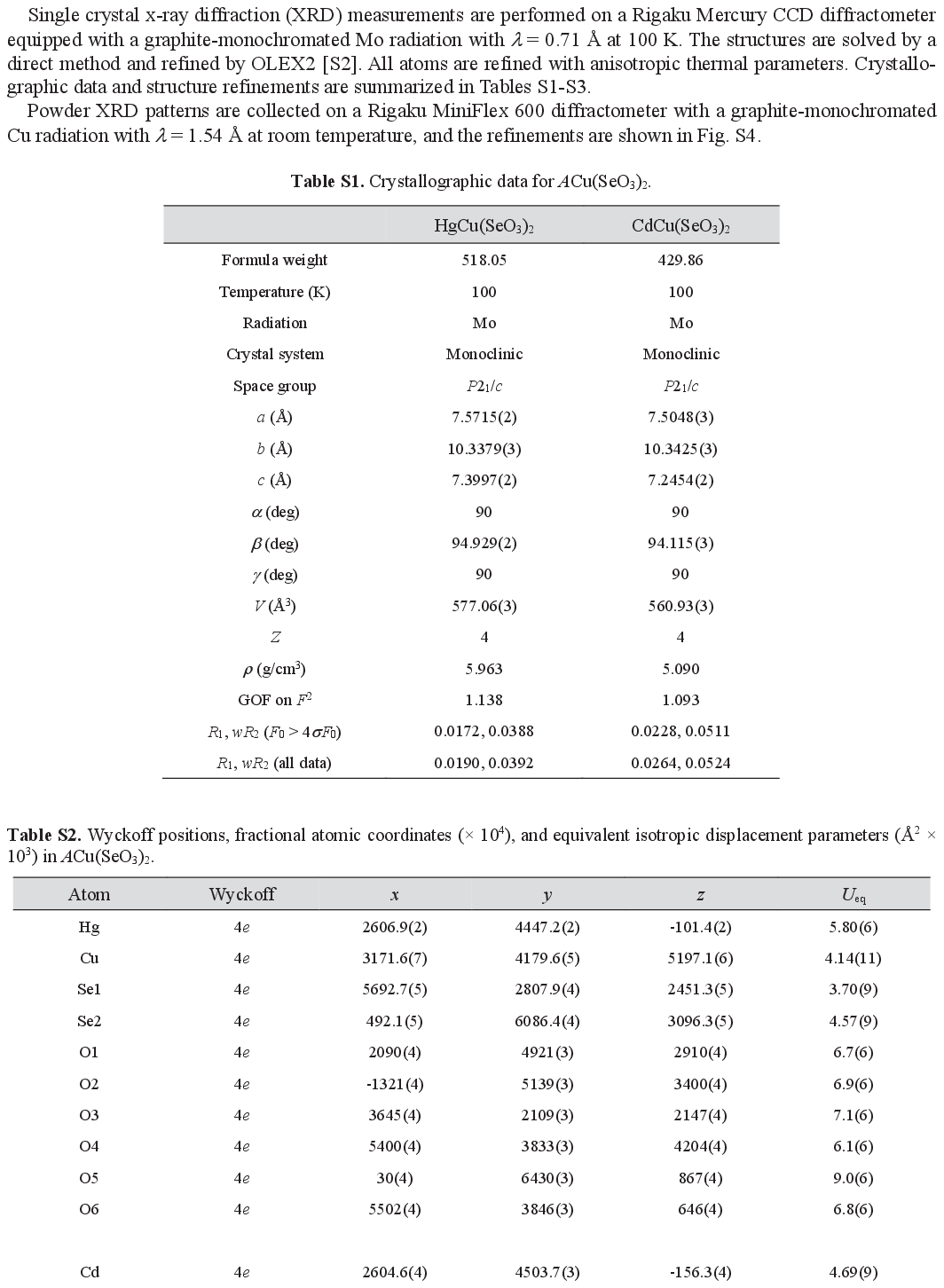}
\end{figure*}

\begin{figure*}
\centering
\includegraphics[clip,width=17.5cm]{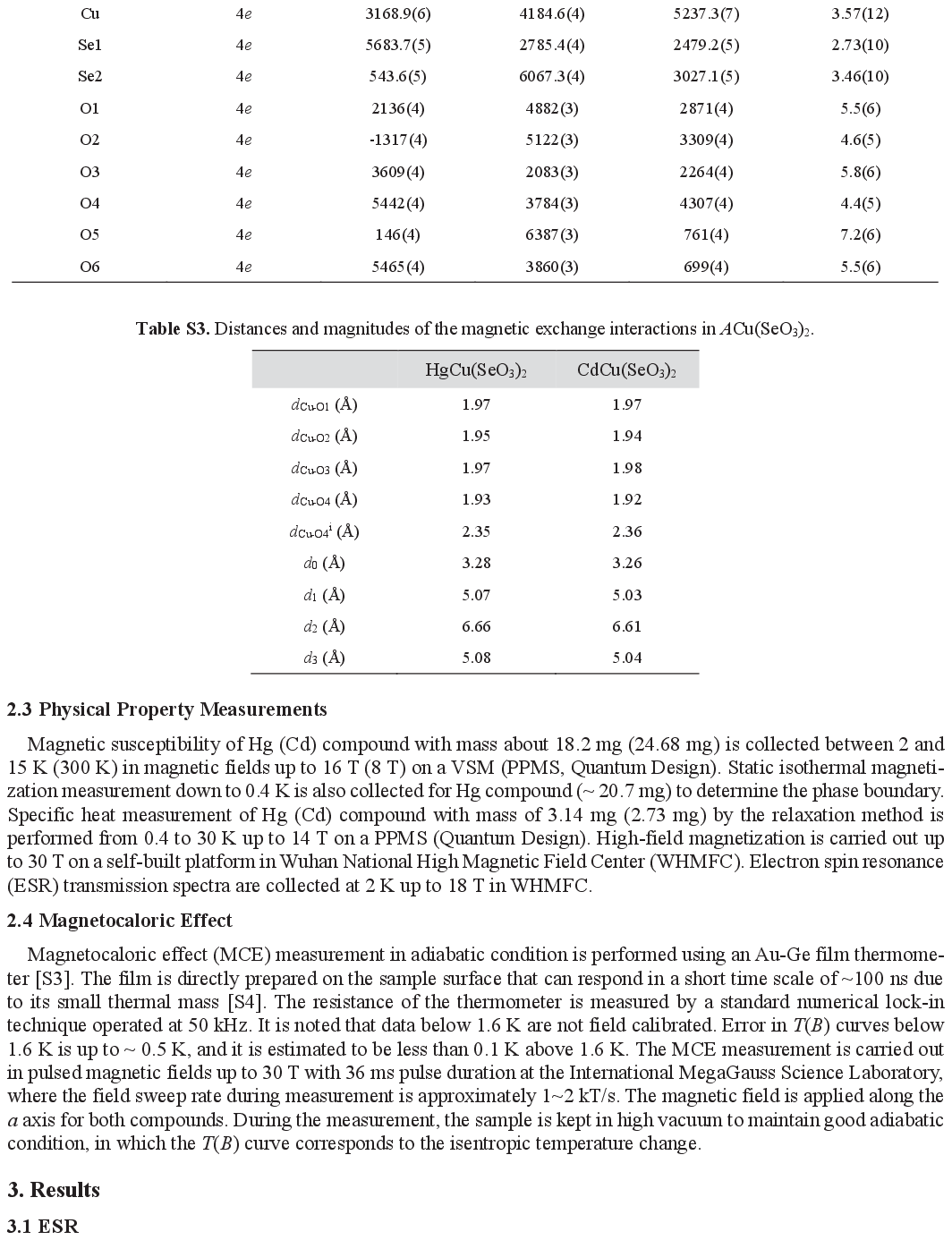}
\end{figure*}

\begin{figure*}
\centering
\includegraphics[clip,width=17.5cm]{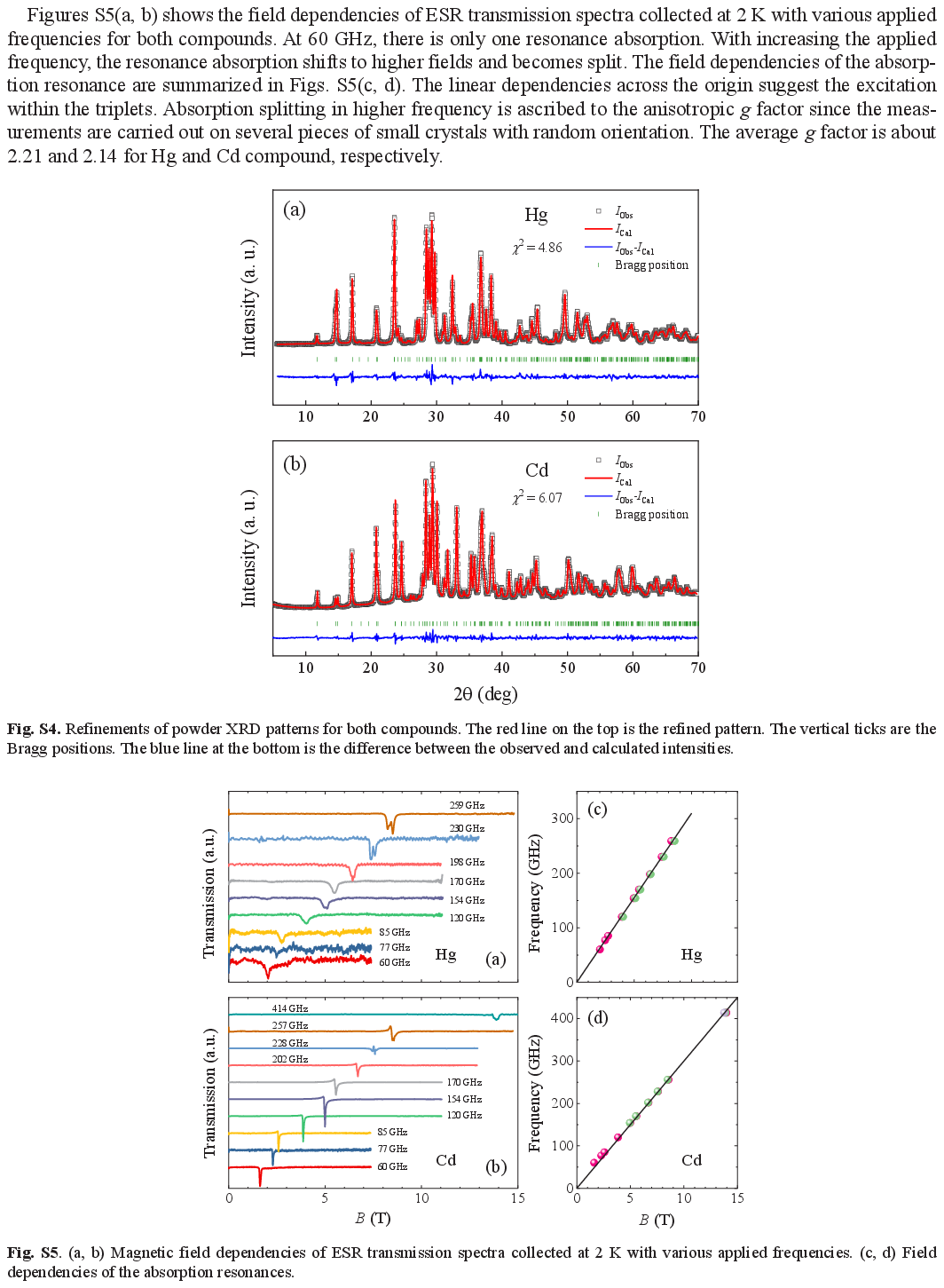}
\end{figure*}

\begin{figure*}
\centering
\includegraphics[clip,width=17.5cm]{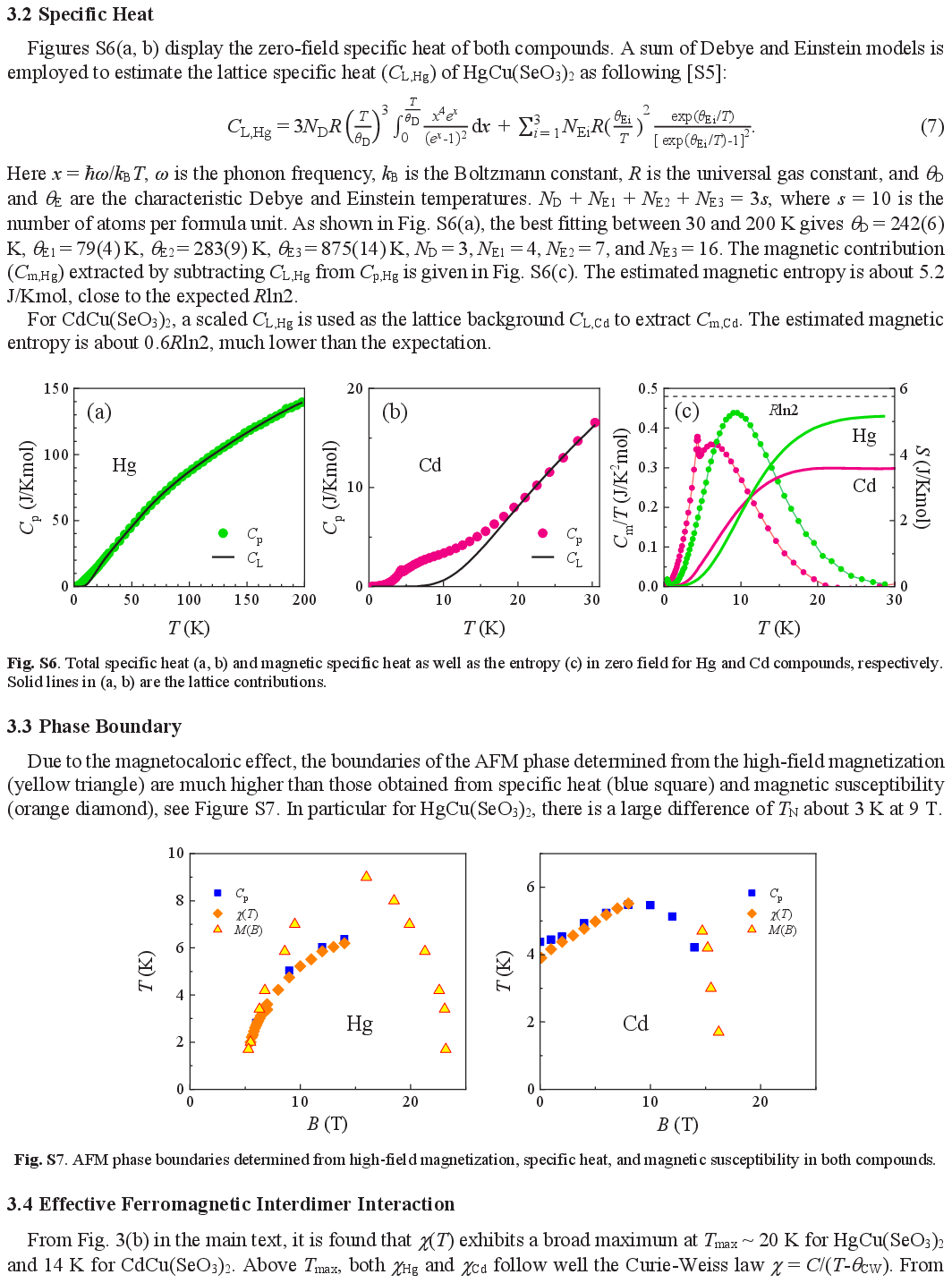}
\end{figure*}

\begin{figure*}
\centering
\includegraphics[clip,width=17.5cm]{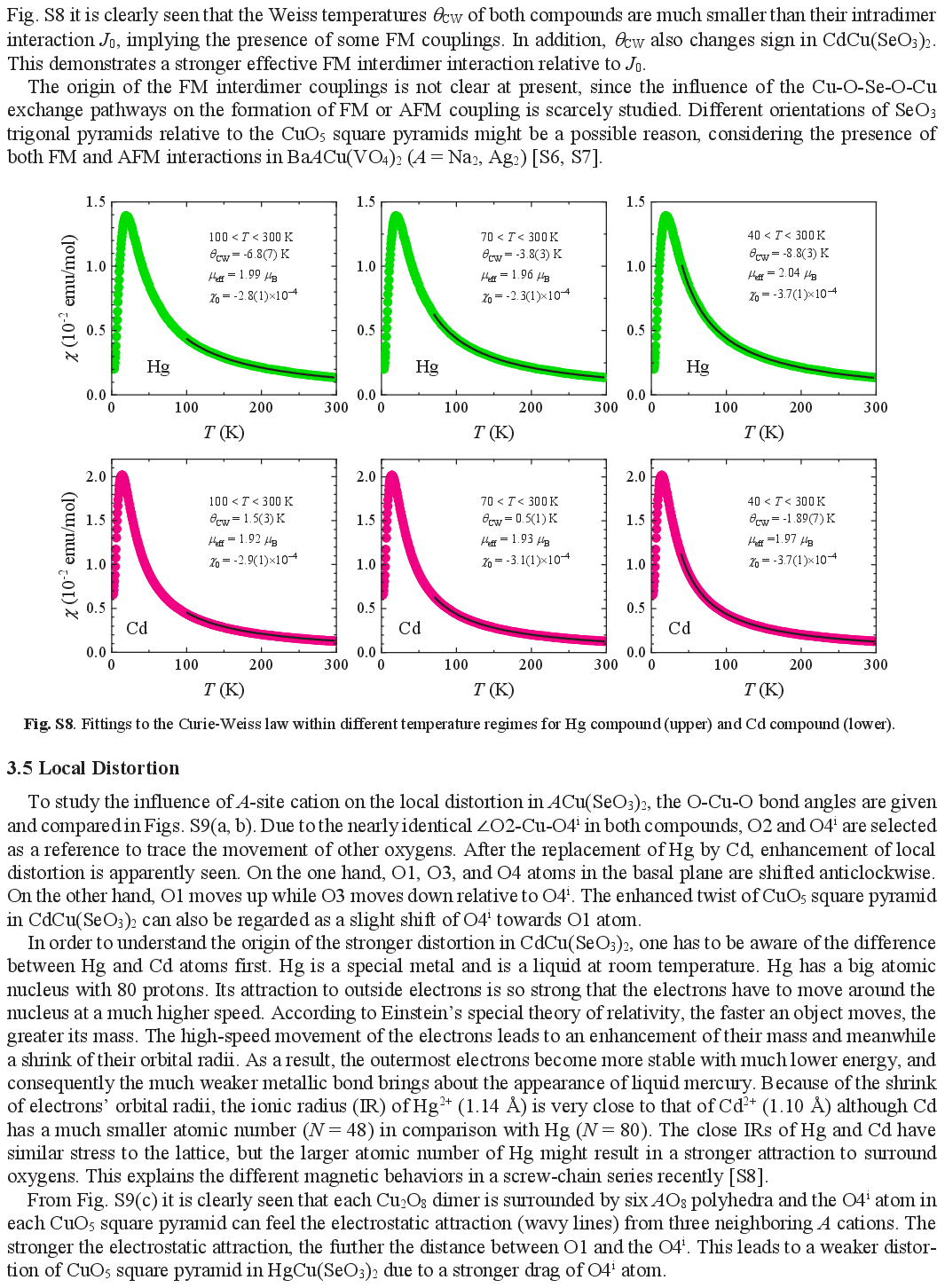}
\end{figure*}

\begin{figure*}
\centering
\includegraphics[clip,width=17.5cm]{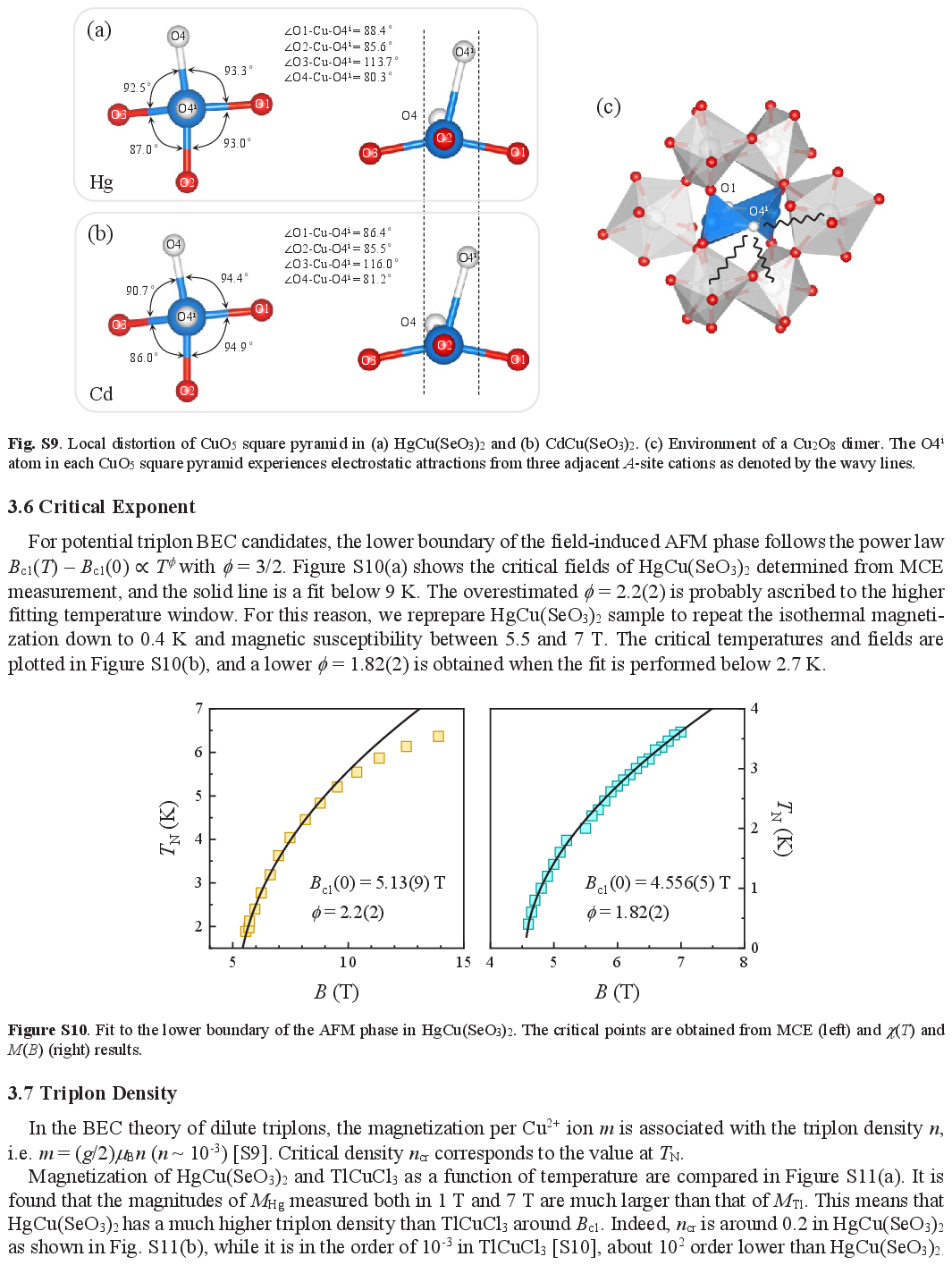}
\end{figure*}

\begin{figure*}
\centering
\includegraphics[clip,width=17.5cm]{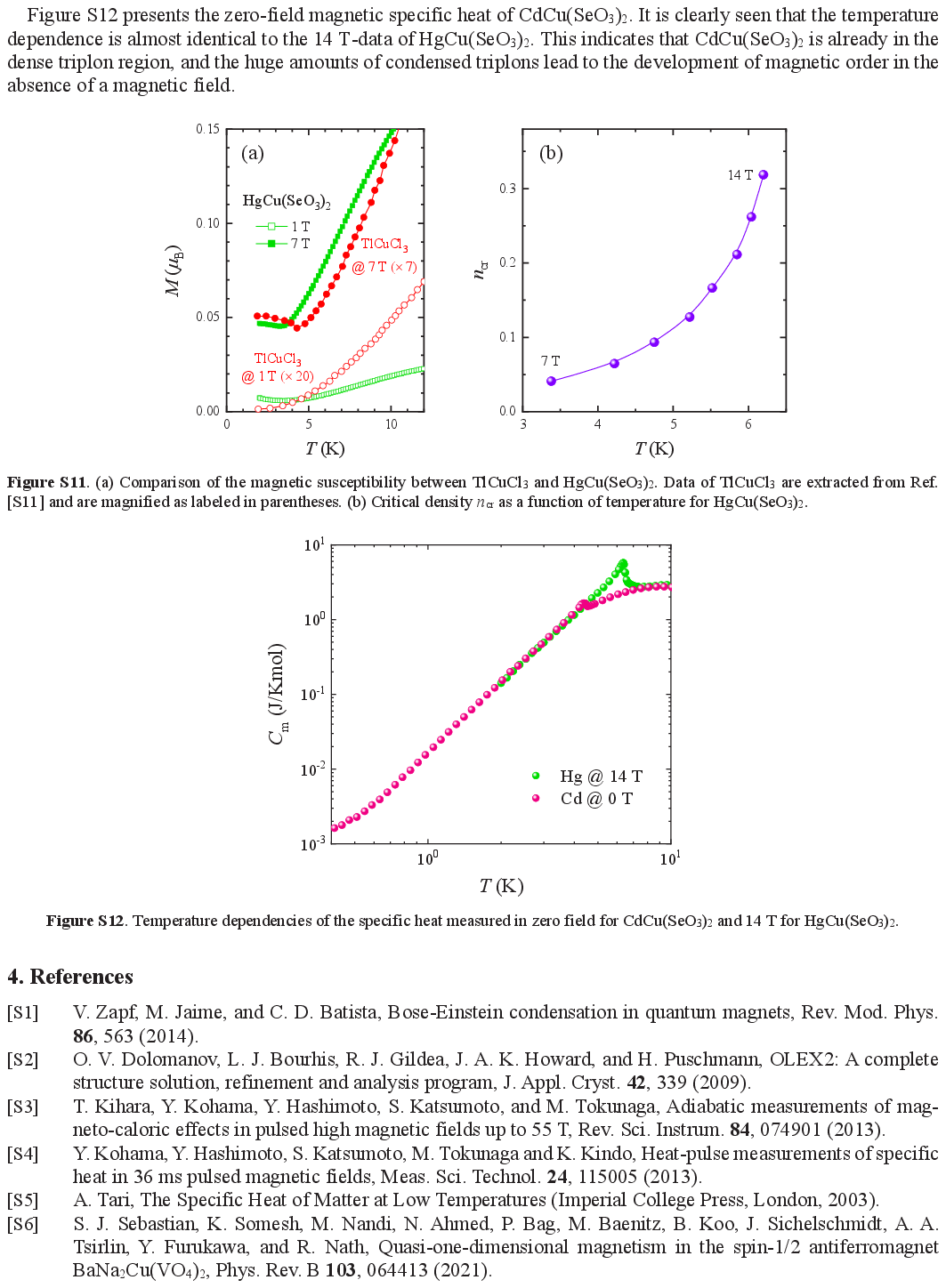}
\end{figure*}

\begin{figure*}
\centering
\includegraphics[clip,width=17.5cm]{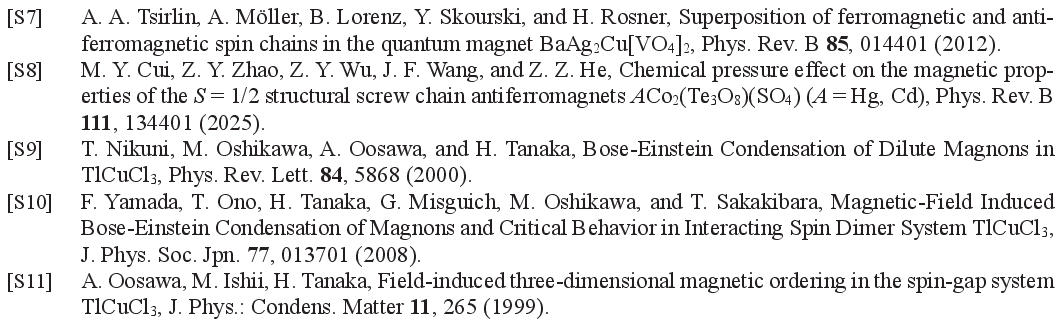}
\end{figure*}

\end{document}